\documentclass[aps,twocolumn]{revtex4}

\usepackage{psfrag,graphicx}
\usepackage{dcolumn}
\usepackage{amsmath,amssymb}
\usepackage{bm}
\usepackage{amsthm}
\theoremstyle{plain}
\newcommand{\identity}{\openone}

\newcommand{\proj}[1]{\mbox{$|#1\rangle \!\langle #1 |$}}
\newcommand{\Tr}{{\text{Tr}}}

\newtheorem{theorem}{Theorem}

\newtheorem{definition}{Definition}
\newcommand{\half}{\mbox{$\textstyle \frac{1}{2}$}}
\newcommand{\ket}[1]{\left | #1\right \rangle}
\newcommand{\bra}[1]{\left \langle #1\right |}

\begin{document}

\title{Multipartite purification protocols: upper and optimal bounds}

\author{Alastair Kay}
\affiliation{Centre for Quantum Computation, Department of Applied
Mathematics and Theoretical Physics, University of Cambridge, Cambridge CB3
0WA, UK}
\author{Jiannis K. Pachos}
\affiliation{Centre for Quantum Computation, Department of Applied
Mathematics and Theoretical Physics, University of Cambridge, Cambridge CB3
0WA, UK}

\date{\today}

\begin{abstract}

A method for producing an upper bound for all multipartite purification
protocols is devised, based on knowing the optimal protocol for purifying
bipartite states. When applied to a range of noise models, both local and
correlated, the optimality of certain protocols can be demonstrated for a
variety of graph and valence bond states.

\end{abstract}

%\pacs{}

\maketitle

\section{Introduction}

Quantum states of many qubits are essential ingredients in the functioning
of quantum computers, and yet their properties, such as entanglement, are
poorly understood.  Of particular interest are graph states, which provide a
range of benefits for communication and
cryptography~\cite{christandl-2005-3788}. In addition, they form a universal
resource for quantum computation~\cite{cluster1,cluster2} and enable
computation in scenarios where the employed two-qubit gate is
probabilistic~\cite{lim-2005-95,kok:04,Jens:06}. However, any practical
implementation will introduce noise to the system. The noise-induced errors
need to be minimized, and corrected, before a practical application of these
states is considered. One way to achieve this is by purification, where many
copies of the noisy state are combined to yield a single perfect copy.

The concentration of entanglement in two-qubit systems has yielded some
important results in quantum information, such as secure communication
via privacy amplification~\cite{bipartite_purification:2}. Nevertheless, the
detailed study of similar transformations in many-qubit systems has only
recently commenced. The purification of two-qubit states was first examined
in~\cite{bipartite_purification:2,bipartite_purification:3}, and the
performance of these protocols is optimal if the operations are perfect. The
efficacy of the proposed protocols in the presence of noise was explored in
\cite{bipartite_purification:1}. Subsequently, the question of purifying
multipartite states has arisen \cite{murao:98}. A variety of protocols have
been discussed, starting from a subset of graph states
\cite{dur:03,aschauer:04} and generalizing to arbitrary graph
\cite{Briegel:06} and stabilizer states~\cite{Knill:06}. These different
protocols tend to trade between a large tolerance of
noise~\cite{dur:03,aschauer:04} and the scaling of purification
rate~\cite{Raussendorf:06}. To date, little has been said on the subject of
what is optimal, although some (non-tight) bounds have previously been
found, such as in the case of independent local
$Z$-noise~\cite{purify_thermal}, or for GHZ
states~\cite{PhysRevLett.83.1054,PhysRevA.59.141,dur:99}.

In this paper, we prove the optimality of certain purification protocols for
a variety of noise models by considering general upper bounds. These
derivations are based on a central theorem that analyzes the purification of
multipartite states in terms of the purification of bipartite states, and
hence allows for a direct extension of previous optimality proofs. When
restricted to $Z$-noise, the application of the theorem becomes
straightforward for a certain class of graph states called locally
reconstructible states. These states allow for the direct application of the
optimal bipartite purification protocol for each link of the graph, thus
extending it to the multipartite case. Subsequently, a wide range of upper
bounds in the tolerated error rates are derived for a variety of error
models and states, while optimality is numerically demonstrated in certain
cases.

The paper is organized as follows. After an initial introduction to graph
states (Sec.~\ref{sec:graphs}) and purification protocols
(Sec.~\ref{sec:purification}), we introduce our main theorem in
Sec.~\ref{sec_optimal}, which proves that if purification of a bipartite
state is impossible, so is the purification of related multipartite states.
This result is applied to a variety of error models in
Sec.~\ref{sec:upperbound}, including local $Z$-noise and depolarizing noise.
For the case of $Z$-noise we prove optimality of the two purification
protocols under consideration for a sub-class of graph states. In the case
of maximally depolarizing noise, we prove a universal bound which applies to
all graph states. Finally, in Sec.~\ref{sec:vbs}, we derive
an example of a valence bond state that can be optimally purified, proving
that our method is not merely limited to graph states. Critically, the
example that we produce has a finite entanglement length. The presented
results expand and extend the work of~\cite{Kay:2006b}.

\section{Graph States} \label{sec:graphs}

For the majority of this paper, we will be interested in the purification of
graph states. These can be defined in two equivalent ways. With a particular
graph $G$, we can associate a set of vertices, $V_G$, and edges, $E_G$, which
connect pairs of vertices. The first way to define a pure graph state is as
the ground state of the Hamiltonian
\begin{equation}
H=\sum_{i\in V_G}J_iX_i\!\!\!\prod_{\{i,j\}\in E_G}\!\!\!Z_j,
\label{eqn:define_graph}
\end{equation}
where we have attached a qubit to each vertex and $X_i$ and $Z_j$ are the
familiar Pauli matrices applied to qubits $i$ and $j$ respectively. The
individual terms $K_i=X_i\prod_{\{i,j\}\in E_G}Z_j$ commute with each other,
$[K_i,K_j]=0$, and thus stabilize the graph state. These are scaled by
arbitrary coupling strengths $J_i$, which we will subsequently take to be
equal. We note that $[K_i,Z_j]=0$ if $i\neq j$, and $\{K_i,Z_i\}=0$, which
means that the excited states of the Hamiltonian are described by local
$Z$-rotations, and that these local rotations constitute a complete,
orthogonal basis over the Hilbert space associated with the graph. An
equivalent way to define a graph state is in a more operational sense, where
we prepare each qubit in the state $\ket{+}=(\ket{0}+\ket{1})/\sqrt{2}$, and
apply controlled-phase gates along each edge of the graph.

For what follows, the action of measurements on a graph state are of particular interest \cite{hein:04}. A $Z$-measurement
removes a vertex from the graph, along with its edges. If we wish to form a
graph state from two-qubit states corresponding to the edges of the graph,
it is simplest to revert to the matrix product state formalism \cite{vbs:1},
where we apply measurements on all of the qubits that need to be combined to
a single qubit. In the case of a linear cluster state, this simply
corresponds to performing a controlled-phase gate between the two pairs,
measuring one qubit in the $X$-basis, and performing a Hadamard rotation on
the other. Although this simple method does not generalize to other graphs,
the examples which we choose to give will be for linear cluster states, and
hence this description is valid. For other graph states, the operation that we
apply must project $N$ qubits on a local site to 1 qubit, and is represented
as
$$P=\ket{0}\bra{0}^{\otimes N}+\ket{1}\bra{1}^{\otimes N}.$$
For example, the required operations to create the pentagon of
Fig.~\ref{fig:twocol}(a) are shown in Fig.~\ref{fig:twocol}(b).

\subsection{Locally Reconstructible States}

\begin{figure}
\begin{center}
\includegraphics[width=0.33\textwidth]{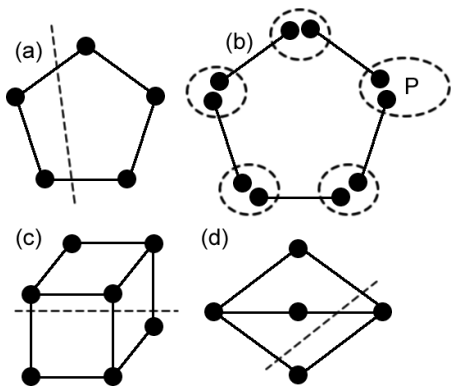}
\end{center}
\caption{(a) A pentagon is locally reconstructible, but not
two-colorable. This graph is locally equivalent to the five-qubit error
correcting code. (b) The pentagon can be formed from two-qubit pairs by
applying the projector $P=\ket{0}\bra{00}+\ket{1}\bra{11}$ to each pair of
qubits that needs to be combined. (c) The graph state which is locally
equivalent to the codewords of the Steane-[[7,1,3]] error correcting code.
(d) A shape which is two-colorable, but not locally reconstructible. }
\label{fig:twocol}
\end{figure}

For a certain error model that we shall address (local $Z$-noise), we will
be particularly interested in the restriction to a class of graph states
which we call Locally Reconstructible (LR), defined as follows:
\begin{definition}
Locally Reconstructible graph states are connected graph states for which there
exists a non-trivial partitioning of the qubits into two parties such that neither party
has more than one edge from each qubit crossing the partition.
\end{definition}

In order to prove that this class of states is non-trivial, we examine some
of its properties in App.~\ref{app:a}. Importantly, this class includes all
cluster states ($d$-dimensional cubes), GHZ states (one vertex with edges to
all others), and graphs which are locally equivalent to the codewords of error-correcting
codes such as the Shor-[[9,1,3]] code, the 5-qubit code (Fig.~\ref{fig:twocol}(a)) and the
Steane-[[7,1,3]] code (Fig.~\ref{fig:twocol}(c)). Indeed, in both of these figures, we provide a partition that demonstrates the LR character of the corresponding graph.

\begin{figure}
\begin{center}
\includegraphics[width=0.35\textwidth]{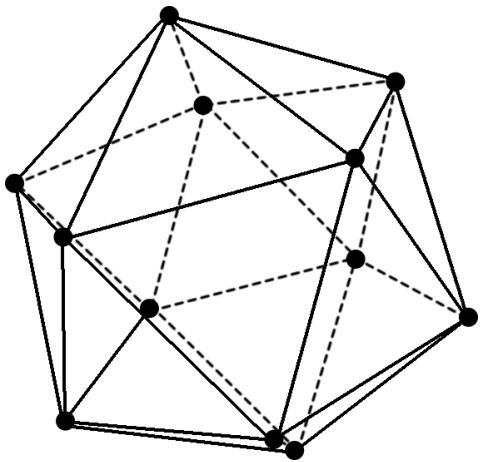}
\end{center}
\caption{The icosahedral graph, which is not equivalent to any LR
states using the local operations $\sqrt{K_i}$.} \label{fig:icos}
\end{figure}

We are not aware of this classification of graph states coinciding with any
previous definition. For example, in Fig.~\ref{fig:twocol} we provide two
examples that show firstly that LR states are not necessarily
two-colorable~\cite{aschauer:04} (Fig.~\ref{fig:twocol}(a)) and, secondly,
that not all two-colorable states are LR (Fig.~\ref{fig:twocol}(d)). In this case, for every possible partition, there is always a qubit that has at least two edges crossing the partition. One such partition is depicted in Fig.~\ref{fig:twocol}(d).
Nevertheless, one can show that for all graphs of up to seven qubits the
corresponding graph states are locally equivalent to LR states. Indeed, all
the graphs in \cite{hein:04}, which are used to categorize the local
unitarily equivalent graphs of up to seven qubits, are LR. That is, one can
reversibly transform pure graph states of seven or fewer qubits to LR states
using local operations $\sqrt{K_i}$. For larger systems, there exist
examples that are not locally equivalent to an LR graph. Specifically, the
icosahedral graph (12 vertices, degree 5, Fig.~\ref{fig:icos}) and its local
equivalences form a collection of 54 graphs, none of which are LR. Since the
operations $\sqrt{K_i}$ are Clifford operators, it remains a possibility
that there are other local unitaries that could be applied which yield a
different set of local equivalences
\cite{hein:04,hein:proc,Zeng:2006}.

%Identifying the sets of local errors which can be optimally purified for a
%given graph is a significant combinatorics problem for which we do not have
%an analytic description. A necessary, but not sufficient, condition is that
%the local errors constitute a complete basis for the graph. This is simply a
%result of the fact that the $Z$-errors on the locally equivalent graph
%constitute a complete basis and the local unitaries cannot change that fact.

%Lacking the tools to describe these local equivalences rigorously, we are
%restricted to testing all possibilities for a given graph. To better examine
%the types of noise for which the proof holds, we choose to examine a
%particular graph, as depicted in Fig.~\ref{fig:steane}. The importance of
%this graph stems from the fact that it is locally equivalent to the
%Steane-[[7,1,3]] code.

%More than $10\%$ (224) of the $3^7=2187$ combinations of local $X$, $Y$ or
%$Z$ errors on this graph can be optimally purified with the same threshold
%fidelity of $1/2^{N/2}$.

\section{Purification Protocols} \label{sec:purification}

The aim of a purification protocol is to take many identical copies of a
noisy state $\rho_G$ and produce a single, pure, copy $\ket{\psi_G}$. We
consider that each qubit in the state $\rho_G$ is held by a different party
(Alice, Bob\ldots), and that they hold the same qubit from every copy. The
parties are restricted to applying Stochastic Local Operations and Classical
Communication (SLOCC), which we initially assume to be perfect. These
restrictions serve to illustrate the entanglement properties of $\rho_G$.
There are also physically motivated systems where locality restrictions come
into play, such as with quantum repeaters
\cite{bipartite_purification:1}.

We assume that $\rho_G$ is diagonal in the graph state basis, i.e.
$$\rho_G=\sum_{j\in\{0,1\}^N}\lambda_jZ_j\proj{\psi_G}Z_j,$$
where $j$ indicates which of the $N$ qubits a $Z$-rotation is applied to. As has previously been
explained elsewhere (see, for example, \cite{aschauer:04}), any non-diagonal state can be made
diagonal by SLOCC without changing the diagonal elements by probabilistically applying the stabilizers (with probabilities suitably chosen to negate the off-diagonal elements). The numbers
$0\leq\lambda_j\leq 1$ encapsulate all the information about the errors.
We are assuming that we know all of these values i.e.~we know the noise model, although it is not {\em strictly} necessary to know the error probability, $p$. Hence, without loss of generality, we assume that $\lambda_0$ is the largest
element, since we wish to purify towards $\ket{\psi_G}$. This is the
fidelity of the initial state $\rho_G$. If $\lambda_0$ were not the largest value, one could apply $Z$-rotations to make it the largest value, as these permute the diagonal elements.

In this paper, we are interested in proving universally applicable bounds.
We shall define the {\em{threshold fidelity}} $f_\text{threshold}$ as the
value of $\lambda_0$ below which purification is impossible, regardless of
the protocol employed, for a specific noise model. All protocols have their
own critical fidelity $f_\text{crit}\geq f_\text{threshold}$ below which
that protocol does not work. Optimality of a protocol is proven when
equality holds.

\subsection{Genuine Multipartite Purification}

In the following subsection, we will propose a protocol that is based on
bipartite purification and thus is easily analyzed. Indeed, for some
types of noise and classes of states, one can prove its optimality. However,
this protocol will not perform well in general, and so when we are interested in other
noise models, we are forced to resort to a multipartite purification
protocol, such as the one described in~\cite{dur:03,aschauer:04}. We refer
to this specific protocol as the Genuine Multipartite Purification Protocol
(GMPP).

The GMPP is applied to two-colorable graph states, which are graph states
where every qubit can be labelled as either A or B such that all edges
connect an A and a B. The protocol proceeds by application of arbitrarily
ordered sequences of two sub-protocols $P1$ and $P2$. Since $Z$-rotations
form a complete basis for the graph state, the state can be labelled by
vectors $\mu_A$ and $\mu_B$, specifying which A or B qubits, respectively,
have $Z$-rotations applied to them. The action of the two protocols is
\begin{eqnarray}
P1:\lambda_{\mu_A,\mu_B}&=&\sum_{\nu_B}\lambda_{\mu_A,\nu_B}
\lambda_{\mu_A,\mu_B\oplus\nu_B}    \nonumber\\
P2:\lambda_{\mu_A,\mu_B}&=&\sum_{\nu_A}\lambda_{\nu_A,\mu_B}
\lambda_{\mu_A\oplus\nu_A,\mu_B}    \nonumber
\end{eqnarray}
which then have to be renormalized. Both are realized by
post-selecting on particular measurement results, which means that the rate
of purification decreases exponentially with the number of qubits present.
The application of the GMPP is challenging to analyze due to the arbitrary
choice of the sub-protocols $P1$ and $P2$ at each step. In general, we
resort to numerical exploration which, with finite computational resources,
can never tell us precisely how close the GMPP comes to any upper-bound.
Another multipartite purification protocol has recently been proposed
\cite{Raussendorf:06}, which is easier to analyze for a range of errors, and
achieves a superior purification rate. However, the critical fidelities at
which it works are larger than for the GMPP, so little benefit can be
derived from comparing them to the threshold fidelities which we calculate.
Other recent work \cite{Knill:06,Briegel:06} has described protocols which
are not limited to two-colorable states.

\subsection{Bipartite-based Purification}

\begin{figure}[!t]
\begin{center}
\includegraphics[width=0.44\textwidth]{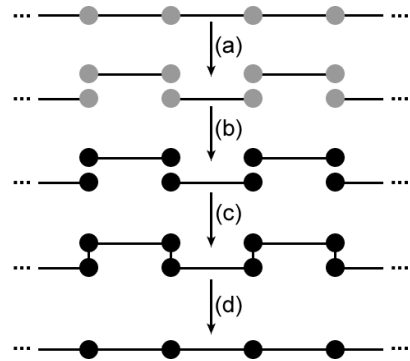}
\end{center}
\caption{The Divide and Rebuild Purification Protocol. We start with
  many copies of the noisy graph state. (a) We form
  two-qubit nearest-neighbor states (noisy) by performing $Z$-measurements.
  (b) Two-qubit states are
  purified (if possible). (c) Controlled-phase gates are applied between
  local qubits. (d) All qubits except one from each party are measured in
  the $X$-basis, leaving the remaining qubits in the purified state.}
\vspace{-0.5cm}
\label{fig:protocol}
\end{figure}

The second purification protocol which we will analyze is certainly not new
(see, for example, \cite{murao:98,dur:04,purify_thermal}), but its
simplicity enables the derivation of rigorous results. To implement the
protocol (Fig.~\ref{fig:protocol}), we initially measure the qubits of the
graph state in the $Z$-basis until we are left with a single two-qubit state
$\rho_2$. Many copies of this state are then used to purify, if possible, a
Bell pair $\ket{\psi_2}$. By performing $Z$-measurements on different sets
of qubits, different Bell pairs are generated. Once we have a Bell pair for
every edge in the graph, we can locally reconstruct the state, e.g.~by
applying controlled-phase gates and $X$-measurements. The conditions under
which
$\rho_2=\text{diag}(\lambda_{00},\lambda_{01},\lambda_{10},\lambda_{11})$
can be purified to $\ket{\psi_2}$ are well-known
\cite{bipartite_purification:2,bipartite_purification:3},
\begin{equation}
\lambda_{00}>\half, \label{eqn:bipartite_condition}
\end{equation}
and have been shown to be optimal using the positive partial transpose
condition \cite{hor:sep,peres:sep}. Hence, we only have to relate
$\lambda_{00}$ to the values of $\lambda_j$ of the original state, $\rho_G$.
We refer to this protocol as the Divide and Rebuild Purification Protocol
(DRPP).

In \cite{Kay:2006b}, we considered the rate of purification for the DRPP,
which is applicable whenever it is the optimal protocol. The rate of
purification, $R_\psi$, for the state $\ket{\psi}$ was described in terms of
the rate of purification of a Bell state, $R_2$, taking the standard
definition of rate,
$$
R_\psi=\frac{\rm{\#\; of\; copies\; of\; } \ket{\psi} \rm{\;
produced}}{\rm{\#\; of \: copies\; of\; } \rho \rm{\; consumed}}.
$$
This allowed us to bound the rate of purification by
$$
R_2\geq R_\psi\geq \frac{R_2}{N_\text{geo}}.
$$
where $N_\text{geo}$ is a small geometric factor determined by the graph.
This rate is a vast improvement over the GMPP, although other protocols have
better rates at the cost of being less robust \cite{Raussendorf:06}. For
$d$-dimensional cluster states, it was shown that the geometrical factor
$N_\text{geo}=3d^2$ is independent of the size of the graph. A similar
argument can be applied to general graph states, and yields an upper bound
$$N_\text{geo}\leq \min\left(2(D_G-1)D_G+1,\binom{N}{2}\right)$$ where $D_G$
is the maximum degree of $G$ i.e.~no vertex has more than $D_G$ edges. This
does not coincide with the result for cluster states because we used
knowledge of the geometry of cluster states to optimize our use of
resources.

\subsection{Upper-Bound to the Purification of Multipartite States}
\label{sec_optimal}

\begin{figure}[!t]
\begin{center}
\includegraphics[width=0.45\textwidth]{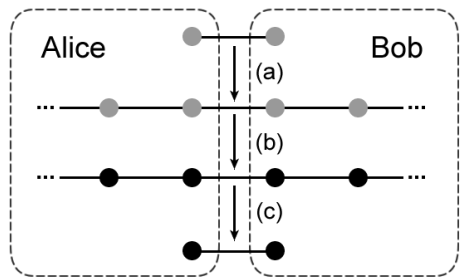}
\end{center}
\caption{If we assume the existence of a purification protocol
for the multipartite state, then this implies that we can purify the
two-qubit state. (a) Alice and Bob take the two-qubit state and reconstruct
the noisy graph state. (b) This state is purified. (c) All extra qubits are
measured out to return the original pair, now pure.}
\label{fig:optimality}
\end{figure}

The intention of this paper is to make statements about when purification is
impossible for all protocols. To provide such a proof, we consider two
parties Alice and Bob, each locally handling many qubits. The operations
they perform are more general, but include, multipartite operations.

\begin{theorem}
\label{main_theorem} Consider the scenario where we wish to purify a
two-qubit state $\rho_2$. Provided that many copies of $\rho_2$ can be
converted by SLOCC into a noisy graph state $\rho_G$ held by the two
parties, then if purification of $\rho_2$ is impossible, so is purification
of $\rho_G$.
\end{theorem}
\begin{proof}
The veracity of this theorem is shown by contradiction. We assume that
purification of $\rho_G$ is possible by some protocol. From the condition of
the theorem, we can start with $\rho_2$ and locally convert it into
$\rho_G$ by employing extra qubits. By assumption, there exists a protocol
that can purify this state, leading to the pure state $\ket{\psi_G}$. This
can be converted into a maximally entangled two-qubit state by projecting
out the additional qubits with local $Z$-measurements. Hence, $\rho_2$ can
always be purified if $\rho_G$ can be purified. If we know that $\rho_2$
cannot be purified, we have a contradiction, and the initial assumption must
be false. These steps are depicted in Fig.~\ref{fig:optimality}.

\end{proof}

A corollary is that if the state $\rho_G$ is described by a parameter $p$
indicating the probability of an error occurring, this theorem gives an
upper-bound on the value of $p$ such that $\rho_G$ can be purified. This
could alternatively be viewed as a lower-bound on the required fidelity of
the state $\rho_G$. We choose to refer to it as an upper-bound.

\begin{figure}[!t]
\begin{center}
\includegraphics[width=0.45\textwidth]{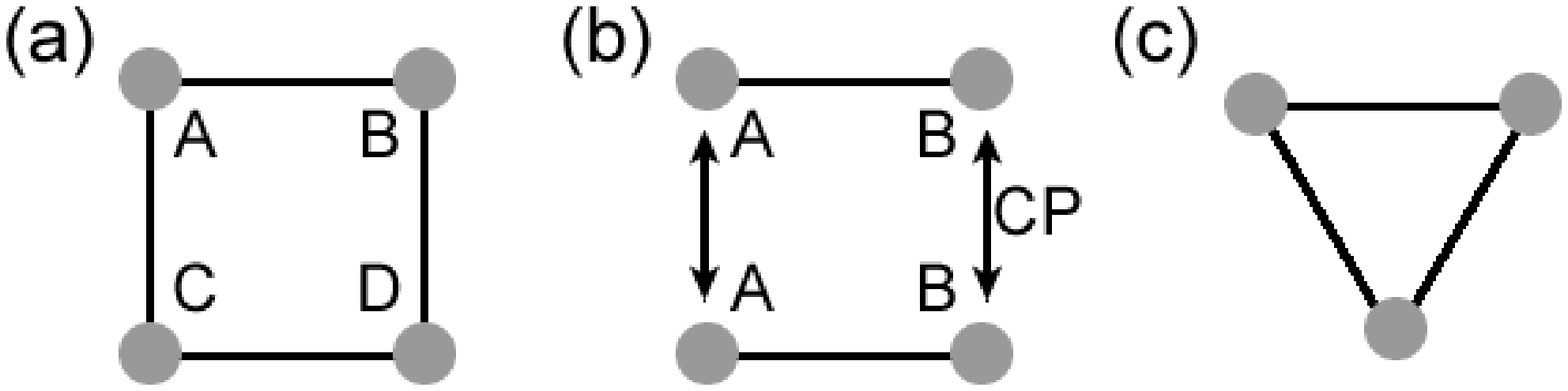}
\end{center}
\caption{(a) Square graph shared between 4 parties. (b) Alice and
Bob can locally reconstruct the square graph using two copies of $\rho_2$,
and applying controlled-phase gates between them. (c) The triangular graph
is the simplest configuration for which the optimality proof fails.}
\label{fig:graphs}
\end{figure}

Little is known about the conversion between bipartite (multi-level) mixed
states $\rho_2^{\otimes n}\rightarrow\rho_G$ \cite{Verstraete:05}, as
required for the local reconstruction condition of Theorem
\ref{main_theorem}. Thus, we have to examine different noise models on a
case-by-case basis, which we do in Sec.~\ref{sec:upperbound}. For LR graphs,
such as Fig.~\ref{fig:graphs}(a), local reconstruction simply involves
replacing each link across the Alice/Bob partition by a single copy of the
two-qubit state. However, for non-LR graphs, such as
Fig.~\ref{fig:graphs}(c), this is not possible because two links need to be
replaced which connect to a single qubit. While reconstruction is still
possible in these cases, it generally means an increase in the local error
probability, which becomes correlated in a different way to the errors in
the rest of the graph. This makes analysis more difficult, weakening
the bounds that one can derive.

Theorem \ref{main_theorem} can be generalized by making two further
observations. Firstly, it is not necessary to restrict to a bipartite state
$\rho_2$, since any state which can be locally converted into the state $\rho_G$
could be used. However, the only existing optimality conditions apply to
two-qubit states. Secondly, the states that we use need not be graph states,
it is just that the formalism of graph states guarantees that we can convert
$\ket{\psi_G}$ into $\ket{\psi_2}$. In the general case, we should be able
to perform measurements on $\ket{\psi_G}$ which return a pure 2-qubit state
with non-zero entanglement, and can subsequently be distilled to a
maximally-entangled state. In Sec.~\ref{sec:vbs}, we apply this method to
the concrete example of a valence bond state.

\subsection{Comparison to the positive partial transpose condition}

Essential to the application of Theorem \ref{main_theorem} is the knowledge
of when a two-qubit state can be purified, which can only happen if there is
non-zero distillable entanglement between the two parties. We can therefore
interpret Theorem \ref{main_theorem} as stating that purification of a
multipartite state is impossible if there is a bipartite split for which
there is no distillable entanglement. Following this interpretation, we can
describe existing multipartite purification protocols as examples of
bipartite purification protocols which use separable operations. In
particular, we observe that the operation of the GMPP is closely related to
the protocol of \cite{alber:01}. The action of this high-dimensional
bipartite purification protocol can be analyzed \cite{delgado:03}, enabling
rigorous comparison between its performance and the upper-bounds calculated
in this paper.

With the aforementioned entanglement interpretation, it is simple to see
that Theorem \ref{main_theorem} is a constructive statement that can
sometimes be applied to discover if a state can be written in
the form
$$
\sum_ip_i\rho_i^A\otimes\rho_i^B.
$$
Consequently, we should expect similar results to those of \cite{dur:99},
where the positive partial transpose (PPT) condition is applied to bipartite
groupings of multipartite states. However, in \cite{dur:99}, extra
depolarization steps are applied which tend to remove entanglement from the
system and, consequently, tight bounds are not expected.

In general, how should our technique compare to the PPT condition, if one does not introduce additional depolarizing steps? As expressed in terms of a reconstruction from two-qubit states, our upper bounds are strictly weaker than PPT. We can see that this is the case because starting from any two-qubit state which is separable (for which the PPT condition is necessary and sufficient), and applying SLOCC results in a state which necessarily has PPT i.e.~is separable, and purification is impossible. However, there also exist states with PPT which cannot be generated from two-qubit separable states, which are known as bound-entangled states. Therefore our condition is strictly weaker than PPT.

Nevertheless, our condition has two major benefits. Firstly, as already indicated, we need not be restricted to reconstructing from two-qubit states. In particular, {\em if} there exist bound entangled states with non-PPT (this still remains an open question), then using these as a basis for reconstruction yields a stronger bound on purification than the PPT condition can provide. Secondly, our technique is constructive, which eases its application in many scenarios, including the situation where the gates applied during the purification procedure are faulty (this will be explored in a later paper).

\section{Upper-Bounds for Various Error Models}
\label{sec:upperbound}

Given Theorem \ref{main_theorem}, it is interesting to apply the method to
different types of noise, yielding bounds on when noisy states are not
purifiable. It may not be possible to attain these bounds with purification
protocols, and we will be able to demonstrate that the DRPP does not always
achieve them. Numerical studies of the GMPP indicate a much tighter match in
performance, although given the asymptotic approach to the bound, and the
existence of strong local attractors, in most cases it is impossible to
precisely verify whether they match.

\subsection{Local $Z$-Noise}
\label{sec:Znoise}

A straightforward application of our theorem comes when considering local
$Z$-noise. While a very restrictive noise model, it has two physical
motivations, namely that the thermal state of the Hamiltonian in
Eqn.~(\ref{eqn:define_graph}) is equivalent to the ground state with local
$Z$-noise, and that $Z$-noise is a significant source of error in some
experimental implementations, such as optical lattices. Moreover, as we
will see, within our treatment this type of noise is the worst-case noise,
giving the lowest probability threshold of all the local noise models considered
here.

We assume that a $Z$-error occurs on each qubit independently, with
probability $p$. When we restrict to LR states, local reconstruction follows
by replacing any links across the bipartition with noisy two-qubit states.
The structure of the class guarantees that the only subsequent operations
that we need to perform are local controlled-phase gates between qubits.
Since these gates commute with $Z$-errors, then starting with a two-qubit
state, a many-qubit state can be built with the same error probability. The
two-qubit state cannot be purified if
$$
(1-p)^2<\half,
$$
so this must hold for all LR states.

Similarly, since $Z$-measurements commute with the $Z$-errors, we can show
that the DRPP can purify the whole state provided
$$
(1-p)^2>\half.
$$
Thus, the protocol is optimal, with a threshold probability of
$p=1-1/\sqrt{2}\approx 0.29$. Equivalently,
\begin{equation}
f_\text{threshold}=f_\text{crit}=(1-p)^N=\frac{1}{2^{N/2}}.
\label{eqn:crit_z}
\end{equation}
This provides a useful benchmark to test other purification protocols, such
as the GMPP (see Sec.~\ref{sec:compare}). Note that the DRPP can purify all
graph states with the same critical fidelity.

Local $Z$-noise also corresponds to the thermal state of the Hamiltonian in
Eqn.~(\ref{eqn:define_graph}), and provided we set $J_i=\Delta/2$, the local
error probability is the same at each site,
$$
p=\frac{e^{-\beta\Delta}}{1+e^{-\beta\Delta}}
$$
where the temperature $T$ is encapsulated in the parameter $\beta=1/(k_BT)$.
Given our proof of optimality, this can be phrased as a critical temperature,
\begin{equation}
T_\text{crit}=\frac{-\Delta}{\ln(\sqrt{2}-1)},
\label{eqn:critical}
\end{equation}
which corresponds to one of the bounds found in \cite{purify_thermal}, as
one expects since the argument of \cite{purify_thermal} also involves the
breaking of cluster states into two-qubit states. As discussed in
\cite{Kay:2006b}, it should be possible to probe this temperature with
existing experimental implementations.

\begin{figure}[!tbh]
\begin{center}
\includegraphics[width=0.47\textwidth]{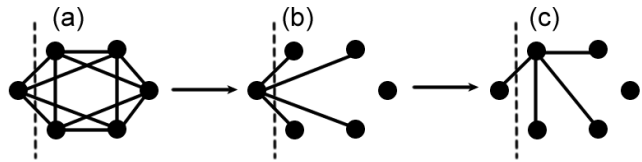}
\end{center}
\caption{(a) Take a general graph and bipartition it with a
single qubit on one side. (b) Apply local controlled-phase gates along all
edges. (c) Apply unitaries $\sqrt{K}$ to change the central node of the GHZ
state. Theorem \ref{main_theorem} can now be applied to obtain upper bounds.}
\label{fig:make_GHZ}
\end{figure}

Thresholds can also be derived for arbitrary graphs by first applying
appropriate transformations. Select from a particular graph, $G$, a single
qubit which has the minimum number of nearest neighbors (minimum degree)
$D_\text{min}$, which we take to constitute a partitioning of the graph. We
now give a reversible procedure which is local to this partitioning to
convert the state into a GHZ state which is LR across the partition. By
applying controlled-phase gates along all edges that are not connected to
our chosen qubit (remember that these gates commute with the $Z$-noise), we
are left with a $(D_\text{min}+1)$-qubit GHZ state as depicted in
Fig.~\ref{fig:make_GHZ}(b). However, the links across the partition are not
the same as for an LR state. To account for this, we apply local unitaries
$\sqrt{K_i}$ on our chosen qubit and one of its neighbors, performing the transformation depicted in
Fig.~\ref{fig:make_GHZ}(c). These local unitaries transform the noise model
from local $Z$-errors to an $X$-error on the original qubit, a $Y$-error on
the center of the GHZ state, and $Z$-errors everywhere else. The
corresponding density matrix can be written in the graph state basis as
\begin{widetext}
\begin{equation}
\rho=\sum_{j\in\{0,1\}^{D_\text{min}-1}}(1-p)^{w_j}p^{w_j}
\text{diag}\left((1-p)^{D_\text{min}+1-2w_j},
p(1-p)^{D_\text{min}-2w_j},p^{D_\text{min}+1-2w_j},
(1-p)p^{D_\text{min}-2w_j}\right)\otimes\proj{j},
\label{eqn:major_state}
\end{equation}
where $w_j$ is the binary weight of $j$. These diagonal elements
$\bra{j}\rho\ket{j}$ are indexed by a number $j\in\{0,1\}^{N-1}$, where the
$i^{th}$ bit indicates whether a $Z$-error has occurred on the $i^{th}$
qubit relative to the desired pure state $\ket{\psi_G}$. The most entangled
pair of qubits is given by $j=0$, so we define
\begin{equation}
\rho_2=\frac{\text{diag}\left((1-p)^{D_\text{min}+1},p(1-p)^{D_\text{min}},
p^{D_\text{min}+1},(1-p)p^{D_\text{min}}\right)}{(1-p)^{D_\text{min}}+
p^{D_\text{min}}}.
\label{eqn:major_state2}
\end{equation}
\end{widetext}
From this state, the entire state $\rho$ in Eqn.~(\ref{eqn:major_state}) can
be reconstructed. Hence, if $\rho_2$ becomes separable, purification must be
impossible, i.e.~if
$$
2(1-p)^{D_\text{min}+1}\leq (1-p)^{D_\text{min}}+p^{D_\text{min}}.
$$
For $D_\text{min}=1$, we recover the bound for the LR graphs as expected.
However, LR graphs need not have $D_\text{min}=1$, proving that this bound
is not always tight. For the triangular and icosahedral graphs, these
calculated threshold probabilities are $p=0.352$ (which the GMPP appears to numerically saturate, thereby exceeding the DRPP) and $p=0.413$ respectively.
As $D_\text{min}\rightarrow\infty$, $p\rightarrow\half$. The bounds given by
\cite{dur:99} in this case also show that purification is impossible if
$p>\half$, but make no tighter claims.

\subsection{Maximally (Global) Depolarizing Noise}

In the previous subsection, we proved optimality of purification for local
$Z$-noise. The local unitary equivalence of graphs can provide similar
bounds for a range of other local noise models. It is now interesting to
examine the case of correlated noise, and to derive bounds in this
context. We will prove bounds for all graph states by initially
restricting to linear cluster states. Choosing the noisy state to be
purified as the maximally depolarized state of an $N$-qubit linear graph state $\ket{\psi_G}$,
$$
\rho_N=\frac{\identity+x\proj{\psi_G}}{2^N+x},
$$
we give an inductive proof which shows how to locally create $\rho_N(x)$
from $\rho_{N-1}(x)$. If we take $\rho_{N-1}(x)$ and add an extra qubit to
it, then the density matrix takes the form
$$
\rho_{N-1}'(x)=\text{diag}(1+x,0,1,0,1,0,\ldots,1,0)/(2^{N-1}+x).
$$
Upon application of a $Z$-rotation to the new qubit (the least significant
bit of $i$), the zeros and non-zeros swap. Further $Z$-rotations on the
other qubits can permute the position of the $1+x$ term. Taking each of
these with probability $(1-p)/2^{N-1}$, or $\rho_{N-1}'(x)$ with probability
$p$, we are left with the density matrix
\begin{eqnarray}
&\frac{p}{2^{N-1}+x}\,\text{diag}\left(1+x,0,1,0,1,0,1\ldots\right)&
\nonumber\\
&+\frac{1-p}{2^{N-1}}\text{diag}\left(0,1,0,1,0,1\ldots\right)&
\nonumber
\end{eqnarray}
This can be forced to take the form of $\rho_N(y)$ by selecting
$$
p=\frac{2^{N-1}+x}{2^N+x}.
$$
and $x=y$. Consequently, the threshold value of $x=2$ for the bipartite case
holds for all $N$, and the threshold fidelity is
\begin{equation}
f_\text{threshold}=\frac{3}{2^N+2}. \label{eqn:max_dep_fid}
\end{equation}
By allowing an arbitrary two-qubit density matrix of the form
$\rho_2=\text{diag}(\half,a,b,\half-a-b)$, it is possible to prove that no
better bound can be given by this method.

In App.~\ref{sec:derivation}, we calculate the performance of the DRPP,
which performs very poorly in the presence of correlated noise, being unable
to purify if $f<1/3$. For $N=2,3$, the GMPP manages to saturate this bound.
However, for $N>3$, we have been unable to find suitable repetitions of $P1$
and $P2$ which purify if $f<1/2^{N/2}$. This is because the
maximally mixed state of one set of errors (i.e.~errors on just the `A'
qubits, with pure `B' qubits), which has fidelity $1/2^{N/2}$, is a strongly
attracting fixed point. In App.~\ref{sec:fixed_derive}, we apply the results
of \cite{delgado:03} to show that for the closely related protocol of
\cite{alber:01} when the qubits are partitioned into two equally sized sets,
$f=1/2^{N/2}$ is indeed a fixed point.

The threshold fidelity for GHZ states is also given by
Eqn.~(\ref{eqn:max_dep_fid}), since the above proof also holds for all
states with a single edge across the bipartite division. The advantage is
that the two potential fixed points in fidelity ($1/2$ and $1/2^{N-1}$) are
different from those of a linear chain. Moreover, the smaller of these is
below the threshold fidelity, and is trivially avoided. For example, for
$N=5$, purification of $x\geq 2.024$ is possible using the GMPP. Given the
anticipated asymptotic approach to the critical fidelity, the GMPP seems to
saturate the bound of Eqn.~(\ref{eqn:max_dep_fid}). The threshold fidelity
for GHZ states coincides precisely with that of \cite{dur:99}, although our
chosen parameterization of the state provides a more convenient condition
for purification.

It is now possible to follow an identical protocol to
Fig.~\ref{fig:make_GHZ} in the case of maximally depolarizing noise to prove
that all graphs are subject to the bound in Eqn.~(\ref{eqn:max_dep_fid}).
This follows trivially from the observation that the application of
controlled-phase gates and local unitaries $\sqrt{K_i}$ do not change the
noise model, only the underlying graph. Once we have a GHZ state with a
suitable partition, the above derivation applies.

\subsection{Local Depolarizing Noise}

\begin{figure}[!t]
\begin{center}
\includegraphics[width=0.48\textwidth]{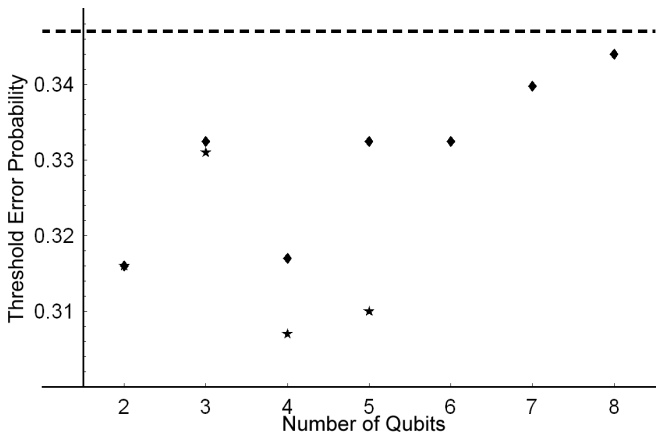}
\end{center}
\caption{Diamonds indicate the probability of a local depolarizing error
above which purification of a linear chain is impossible. The dashed line
indicates a bound below which all such probabilities must lie. Stars
indicate values for which purification can be achieved with the GMPP.}
\label{fig:linear_bounds}
\end{figure}

In addition to local $Z$-noise, a range of other local noise models could be
considered. One such model is where the type of local unitary that is
applied is not known, but it occurs with probability $p$. This is equivalent
to local depolarizing noise occurring with a probability $4p/3$,
\begin{eqnarray}
\mathcal{E}_p^i(\rho)&=&(1-p)\rho+\frac{p}{3}\left(X_i\rho X_i+Y_i\rho
Y_i+Z_i\rho Z_i\right)  \nonumber\\
&=&\left(1-\frac{4p}{3}\right)\rho+\frac{4p}{3}\half
\identity_i\otimes\Tr_i(\rho).  \nonumber
\end{eqnarray}
To derive a threshold, it is possible to follow a similar process to the
previous subsection. Of critical importance to calculating a tight bound is
the optimal selection of the two-qubit state to use across the partition. When considering linear graphs, this state varies with the length of the chain. To illustrate this, we shall
discuss the case of 3 qubits in more detail. The target density matrix, written in the graph state basis, takes
the form
$$
\rho=\text{diag}(a,b,c,b,b,b,b,b),
$$
where $a$, $b$ and $c$ are specified in terms of $p$.
This can be divided into a probabilistic mixture of two components,
%\begin{eqnarray}
%\rho&=&\text{diag}(a,0,c,0,b,0,b,0) \nonumber\\
%&+&\text{diag}(0,b,0,b,0,b,0,b).   \nonumber
%\end{eqnarray}
\begin{eqnarray}
\rho&=&\text{diag}(a,0,c,0,b,0,b,0)+\text{diag}(0,b,0,b,0,b,0,b)
\nonumber\\
&=&\text{diag}(a,c,b,b)\otimes\proj{0}+
\frac{b}{4}\text{diag}(1,1,1,1)\otimes\proj{1}.   \nonumber
\end{eqnarray}
The second of these is a maximally mixed state on the two qubits that have
an edge crossing the bipartite partition, perfectly connected to a third
qubit (with a $Z$-rotation). It is always possible to prepare this, so our
only condition relates to the creation of the first term, which is a
two-qubit state $\text{diag}(a,c,b,b)/(a+2b+c)$ perfectly connected to the
third qubit. This is the form of two-qubit state that we choose to use, and
purification is impossible if $a/(a+2b+c)\leq\half$. Upon evaluation of $a$,
$b$ and $c$, this gives
$$
27-126p+156p^2-64p^3\leq 0,
$$
from which we find the threshold probability of $0.332$. A similar process
can be adopted for all linear graphs, and the results are presented in
Fig.~\ref{fig:linear_bounds}. In App.~\ref{sec:derivation}, we derive the
performance of the DRPP, which fails to achieve these bounds. For 2
qubits, the GMPP is the same as the protocol in
\cite{bipartite_purification:2}, and therefore achieves the two-qubit bound
of $31.7\%$ asymptotically. For 3 qubits, the GMPP comes close to matching
these bounds, purifying at $33.1\%$. However, for $N\geq 4$, the strongly
attracting fixed point of $f=1/2^{\lceil N/2\rceil}$ occurs close to the
threshold probability. This makes numerical analysis particularly
challenging in these cases.

These bounds exceed the bound for local $Z$-noise because for the graph
states, $Z$-noise forms a complete basis, whereas the other errors do not.
Consequently, when two errors coincide, they might cancel, and hence it may
become slightly easier to purify the state. For example, the purification
condition for two qubits subject to local depolarizing noise is
$(1-p)^2+p^2/3>\half$ instead of $(1-p)^2>\half$ ($Z$-noise), where the
extra term comes from cancellation of coinciding errors.

A general bound for a chain of arbitrary length can be obtained by demanding
the conditions under which we can create a particular $N$-qubit state, where
only the first $N-1$ qubits are noisy. Provided $N\geq 3$, we can create a
chain of arbitrary length $M\geq N$ by adding $M-N$ qubits to the end of the
chain, along with the required noise. This bound must be non-increasing with
increasing $N$. Selecting $N=10$, we find that purification is impossible
for all chains of ten or more qubits if $p>0.347$. Consequently, the
threshold probability must tend towards a constant, as observed numerically
for the GMPP in \cite{aschauer:04}. The optimal protocol must also tend to a
constant because of the constant lower-bound provided by the DRPP (App.
\ref{sec:derivation}).

In comparison to linear or GHZ states, the case of the Steane [[7,1,3]] code
is slightly more involved because we need three two-qubit states to cross
the bipartite split. However, all three become separable at the same
threshold probability of $p=0.403$. Numerically, we find that the GMPP
becomes trapped in the fixed point of $f=1/2^4$.

\subsection{Analysis of the GMPP} \label{sec:compare}

\begin{figure}[!t]
\begin{center}
\includegraphics[width=0.33\textwidth]{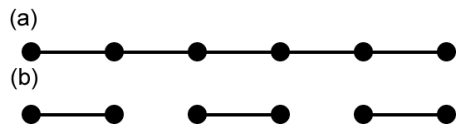}
\end{center}
\caption{For the GMPP, purification is geometry independent.
This means that the purification regimes for the two depicted states are
identical.}
\label{fig:demo}
\end{figure}

To date, evaluation of the performance of the GMPP has resisted analytic techniques, and has instead relied on numerical evaluation, as we have used in this section. While we have observed a close relationship between the GMPP and the analytic bipartite purification protocol of \cite{alber:01,delgado:03}, the direct proof of any connection still remains an open problem. However, we are able to make progress in proving the purification regime for the GMPP in certain special cases. Given this, it becomes interesting to compare the performance of the two purification protocols, the DRPP and the GMPP. Earlier in this section, we have
provided several examples where the GMPP out-performs the DRPP. In this
subsection we will present our analysis of the GMPP, and construct an example in which the GMPP is
provably sub-optimal, being out-performed by the DRPP. This yields an
interpretation as to why the GMPP can get trapped in fixed points for
certain error models.

We proceed by realizing that the sub-protocols of the GMPP depend only on
the diagonal elements of the density matrix, and not the underlying
geometry, other than the numbers of qubits of each color in the
two-colorable graph. This means that the purification regimes of the graphs
depicted in Fig.~\ref{fig:demo} are identical. Hence, one can restrict to
considering purification of the graph in Fig.~\ref{fig:demo}(b). Further, if
we assume that the noise is not correlated between the pairs, and that it is
identical for each pair, then the GMPP is exactly the same as the DRPP,
except that purification of the pairs occurs in parallel instead of
independently, leading to the observed reduction in purification rate. Given
that we can derive the performance of the DRPP, we can deduce the
performance of the GMPP in this case and, consequently, in the case of more
complex graphs such as Fig.~\ref{fig:demo}(a). One such example of noise is
local $Z$-noise, instantly proving that the GMPP is optimal for local
$Z$-noise on LR graphs.

The geometry independence of the GMPP means that it can purify all
two-colorable graphs with local $Z$-noise with the same critical
probability. This includes graphs such as Fig.~\ref{fig:twocol}(d), for
which our analysis gives a threshold probability of $p=0.352$. We interpret
the failure to saturate this bound as the geometry independence of the GMPP
causing it to become trapped by local fixed points.

\begin{figure}[!t]
\begin{center}
\includegraphics[width=0.22\textwidth]{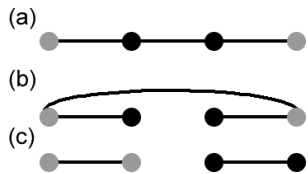}
\end{center}
\caption{Consider purification of the graph in (a) where gray circles
denote qubits with $Z$-errors with probability $p$, and black circles are
pure qubits. For the GMPP, purification of parts (a), (b) and (c) are
identical. (b) provides an upper bound of $p=30\%$ for the GMPP, and (c)
proves that this can be achieved. The DRPP can always purify (a) for
arbitrary $p$.}
\label{fig:suboptimal}
\end{figure}

In Fig.~\ref{fig:suboptimal}, we consider purification of a linear graph
where two of the qubits have a $Z$-error with probability $p$, and the other
two qubits are pure. Given the geometry independence of the GMPP, its
critical probability is the same as for Fig.~\ref{fig:suboptimal}(b). As we
have seen in Sec.~\ref{sec:Znoise}, purification of this graph must be
impossible if $p\geq 1-1/\sqrt{2}$. Consequently, the GMPP cannot purify
Fig.~\ref{fig:suboptimal}(a) if $p\geq 1-1/\sqrt{2}$. A similar manipulation
to Fig.~\ref{fig:demo} yields Fig.~\ref{fig:suboptimal}(c), which can also
be used to show that purification below this critical probability is
possible (application of purification to the pure pair leaves it pure, so we
only have to purify the noisy pair, for which the GMPP is the same as the
optimal two-qubit protocol).

Now consider applying the DRPP to the original chain
(Fig.~\ref{fig:suboptimal}(a)). Each of the three edges to be purified can
always be purified -- one is already pure and the other two only have two
diagonal elements to their density matrices, so can always be purified.
Hence, the state in Fig.~\ref{fig:suboptimal}(a) can always be purified.
This proves that the GMPP is sub-optimal and that in some circumstances the
DRPP out-performs it.

\begin{figure}[!t]
\begin{center}
\includegraphics[width=0.36\textwidth]{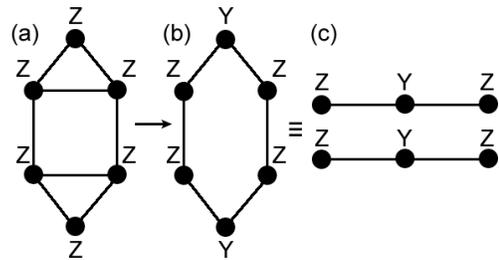}
\end{center}
\caption{Local operations can convert local $Z$-noise on graph (a), which is LR, into a two-colorable graph (b). The GMPP can purify this if it can purify (c). However, this is not the same constraint as the ability to purify a single copy of the three-qubit state.}
\label{fig:pitfall}
\end{figure}

There are certain pitfalls associated with this analysis of the GMPP that we will illustrate with an example. Consider the purification of the triangular graph with local $Z$-noise, for which our upper bound has predicted the impossibility of purification if $p>0.352$. The state can be transformed into a linear graph by local operations where the errors are now $Z,Y,Z$, and since this graph is two-colorable, the GMPP can be applied. Numerically, it appears to saturate our bound. We can now consider the purification of the LR state depicted in Fig.~\ref{fig:pitfall}(a). With local $Z$-noise, we already know that purification is impossible if $p>0.3$. However, we can apply local operations to form a two-colorable state, and subsequently apply the geometry independence of the GMPP to see that purification of this state is equal to the parallel purification of two of the triangular graphs, and hence it appears that it should have the same purification regime. Clearly, there is a discrepancy. This is resolved by observing that the performance of the purification in parallel is not identical to two independent purifications in this case because there is an asymmetry between the $P1$ and $P2$ protocols i.e.~the parallel application requires $P1'=P1\otimes P1$ and $P2'=P2\otimes P2$, where $P1$ and $P2$ are the protocols on the triagle, whereas the GMPP applies $P1'=P1\otimes P2$ and $P2'=P2\otimes P1$.

\section{Purification of Valence Bond States} \label{sec:vbs}

Using a combination of the DRPP and Theorem \ref{main_theorem}, we have
proved that for local $Z$-noise, and some other types of local noise, there
is a threshold error probability below which purification is possible, and
above which purification of graph states is impossible. However, there is no
need to restrict to graph states. By making use of the valence bond
formalism, we will now construct an example of a state which is not a graph
state, but can be optimally purified by the DRPP.

A general valence bond state \cite{cirac:04,vbs:1,DMRG_period,AKLT} can be
described by employing a $D$-dimensional maximally entangled state between
each nearest-neighbor of a graph. Each local party then projects down to a
$d$-dimensional system with a specific projector. If we allow $D=2^{N/2}$,
then any $N$-qubit state can be described by this formalism \cite{cirac:04}.
The class of translationally invariant states can be described
efficiently, using a fixed $D$. These $D$-dimensional maximally entangled
states can be formed from $\log_2(D)$ Bell states. We are solely interested
in constructing a simple example to demonstrate the general properties. As
such, we shall restrict to $D=d=2$ and to a linear graph of 3 qubits. This
contains all the essential properties of valence bond states and,
consequently, we expect that generalizations will follow in a
straightforward manner.

The purification protocol follows the concept of the DRPP, as already
outlined. Our initial state is described by two maximally entangled states
$\ket{\phi}$, joined by a single projector $P_0$, acting on qubits $2$ and
$3$.
$$
\ket{\psi_\text{initial}}=\identity\otimes P_0\otimes\identity
\ket{\phi}_{12}\ket{\phi}_{34}
$$
If local $Z$-noise affects our state, then we apply $Z$-measurements to all
the qubits apart from a single pair, which should retain some entanglement.
Since the noise commutes with the measurement, it suffices to describe what
happens to the pure state,
\begin{eqnarray}
\ket{\psi_2}\ket{0}&=&\frac{1}{2}\identity\otimes P_0\otimes\proj{0}
\left(\sum_i\ket{i}\ket{i}\right)^{\otimes 2}   \nonumber\\
&=&\frac{1}{2}\identity\otimes P_0\otimes\identity\left(\sum_i\ket{i}
\ket{i}\right)\ket{0}\ket{0},  \nonumber
\end{eqnarray}
where we have assumed outcome $\proj{0}$ from the $Z$ measurement. Using
many copies, the state $\ket{\psi_2}$ with local $Z$-noise can be purified
to a two-qubit maximally entangled state, $\ket{\phi}$. We repeat this for
each edge of the graph, and the pure state $\ket{\psi_\text{initial}}$ is
recovered by applying local projectors $P_0$ to each vertex. We can express
the projector $P_0$ as
$$
P_0=\sum_{i,j,k}\alpha_{j,ik}\ket{j}\bra{i}\bra{k}
$$
so that $\alpha^k$ constitutes a $d\times d$ matrix,
$\bra{j}\alpha^k\ket{i}=\alpha_{j,ik}$. Note that this definition does not
coincide with the standard matrix product state definition of these matrices
(where $\bra{j}\alpha^k\ket{i}=\alpha_{k,ji}$).

\begin{figure}[!t]
\begin{center}
\includegraphics[width=0.15\textwidth]{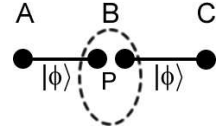}
\end{center}
\caption{The valence bond state that we wish to purify is generated
from two maximally entangled states $\ket{\phi}$, and a projector $P$
applied between them.}
\label{fig:vbs}
\end{figure}

For the optimality proof, we need to start with the state $\ket{\psi_2}$ and
show how to reconstruct $\ket{\psi_\text{initial}}$. We do this by
locally introducing a maximally entangled state $\ket{\phi}$, and applying a
projector $P_1$.
\begin{equation}
\identity\otimes P_1\otimes\identity\ket{\psi_2}\ket{\phi}=
\ket{\psi_\text{initial}}.    \label{eqn:cond1}
\end{equation}
If $Z$-noise is present on $\ket{\psi_2}$, then it must reappear on
$\ket{\psi_\text{initial}}$ when we apply $P_1$,
\begin{equation}
(\identity\otimes P_1\otimes\identity)\cdot(\identity\otimes
Z\otimes\identity\otimes\identity)\ket{\psi_2}\ket{\phi}=
\identity\otimes Z\otimes\identity\ket{\psi_\text{initial}}.
\label{eqn:cond2}
\end{equation}
$P_1$ can be described analogously to $P_0$,
$$
P_1=\sum_{i,j,k}\beta_{j,ik}\ket{j}\bra{i}\bra{k}
$$
which, through Eqn.~(\ref{eqn:cond1}), allows us to show that
$$
\beta^i\alpha^0=\alpha^i.
$$
Subsequent expansion of Eqn.~(\ref{eqn:cond2}) enables the derivation of
a simple condition for when the reconstruction can be performed, and hence
when the optimality proof holds,
\begin{equation}
[\beta^i,Z]\alpha^0=0.  \label{eqn:cond3}
\end{equation}
This does not hold for all valence bond states, but we can construct
examples when it does. In the case where $\alpha^0$ and $\alpha^1$ are
invertible, we find that $\alpha^0(\alpha^1)^{-1}$ must be diagonal.
Applying the optimality proof on both edges of the graph provides a symmetry
between the elements $\alpha_{j,ik}$ and $\alpha_{j,ki}$. This leads to a
final form of the projector
$$
P_0=\left(\begin{array}{cccc}
\frac{\alpha_{0,01}\alpha_{0,10}}{\alpha_{0,11}} & \alpha_{0,01} &
\alpha_{0,10} & \alpha_{0,11}    \\
\frac{\alpha_{1,01}\alpha_{1,10}}{\alpha_{1,11}} & \alpha_{1,01} &
\alpha_{1,10} & \alpha_{1,11}
\end{array}\right).
$$
In the special case of $\alpha_{1,11}=e^{i(\theta_1+\theta_2)}$,
$\alpha_{1,10}=e^{i\theta_2}$, $\alpha_{1,01}=e^{i\theta_1}$ and
$\alpha_{0,11}=\alpha_{0,10}=\alpha_{0,01}=1$, we recover the weighted graph
states \cite{weighted_graphs,hein:proc} and the cluster state
($\theta_1=\theta_2=\pi$). Weighted graph states are identical to the graph
states that have been discussed so far except that to construct the state,
instead of a controlled-$Z$ gate between nearest-neighbors initially in
$\ket{+}$, a controlled-phase gate of arbitrary phase is used. This means
that all the relevant actions continue to commute with $Z$-errors, and we
recover trivially the previous optimality proof, providing a useful
verification of these results. Since the weighted graph states have an
exponentially decreasing localizable entanglement length
\cite{loc_ent,cirac:04}, the general solutions, as described by $P_0$, are expected to have finite localizable entanglement length.

\section{Conclusions}

In this paper, we have described a method which proves that certain noisy
multipartite states cannot be purified. For the case of LR states subject to
$Z$-noise with probability $p$, we have been able to show that for $p>30\%$,
purification is impossible, and that all other states can be purified
i.e.~we have demonstrated optimality of the purification protocol.

Numerical evidence indicates that the GMPP \cite{dur:03,aschauer:04} is
optimal (in terms of the states that can be purified) for a large range of
different errors, including local errors such as $Z$-noise (for which we
have {\em proven} optimality), and non-local errors such as maximally
depolarizing noise. There are cases where the GMPP is not optimal and in all
such cases, we have observed that the protocol gets trapped by strong local
attractors with fidelities $\frac{1}{2^n}$ and $\frac{1}{2^m}$, where the
two-colorable state has $n$ qubits of one color, and $m$ qubits of the
other. These can often be interpreted as being due to the geometry
independence of the GMPP. This also coincides with the results that can be
derived for the bipartite purification protocol (applied to systems of
arbitrary dimension) in \cite{alber:01,delgado:03}. The purification of the
noisy graph in Fig.~\ref{fig:suboptimal}(a) provides proof that
this is a real phenomenon, and not just an artefact of finite computational
resources -- the GMPP has a critical probability for purification, but other
protocols can always purify the graph.

Not only have we demonstrated the optimality of purification of some graph
states, but have also provided an example of a valence bond state which can
be optimally purified, thus demonstrating the general utility of our method.

In forthcoming work, we shall examine how our upper-bound method can be applied to the situation where the gates used during the purification are also faulty, which is a major advantage of our constructive approach. This has potential implications for upper bounds on fault-tolerant thresholds. Interesting extensions of this work could involve taking what we have learnt about the performance of the GMPP and trying to improve it. In particular, we have demonstrated that one should take into account both the geometry of the state and the noise model when constructing a purification protocol, not just the noise model (in the case of the GMPP) or the geometry (in the case of the DRPP). A combined approach, based on the stabilizers of the state, but allowing asymmetry between the terms, appears to be the most sensible approach.

\acknowledgments

We would like to thank Hans Briegel, Wolfgang D\"ur, Robert Raussendorf and
Peter Rohde for helpful conversations. This work was supported by EPSRC,
Clare College, Cambridge and the Royal Society.

\appendix
\section{Analysis of LR Class} \label{app:a}

Given our rather abstract definition of the LR class, it is a worthwhile
exercise to verify that it is a non-trivial class.

\subsection{Graphs of Maximum Degree 3}

Let us consider all possible graphs which have a maximum number of nearest
neighbors (degree) equal to 3 or less. If one of the qubits in the graph has
degree 1, then there exists a trivial partition that shows that the state is
LR.
\begin{center}
\includegraphics[width=0.18\textwidth]{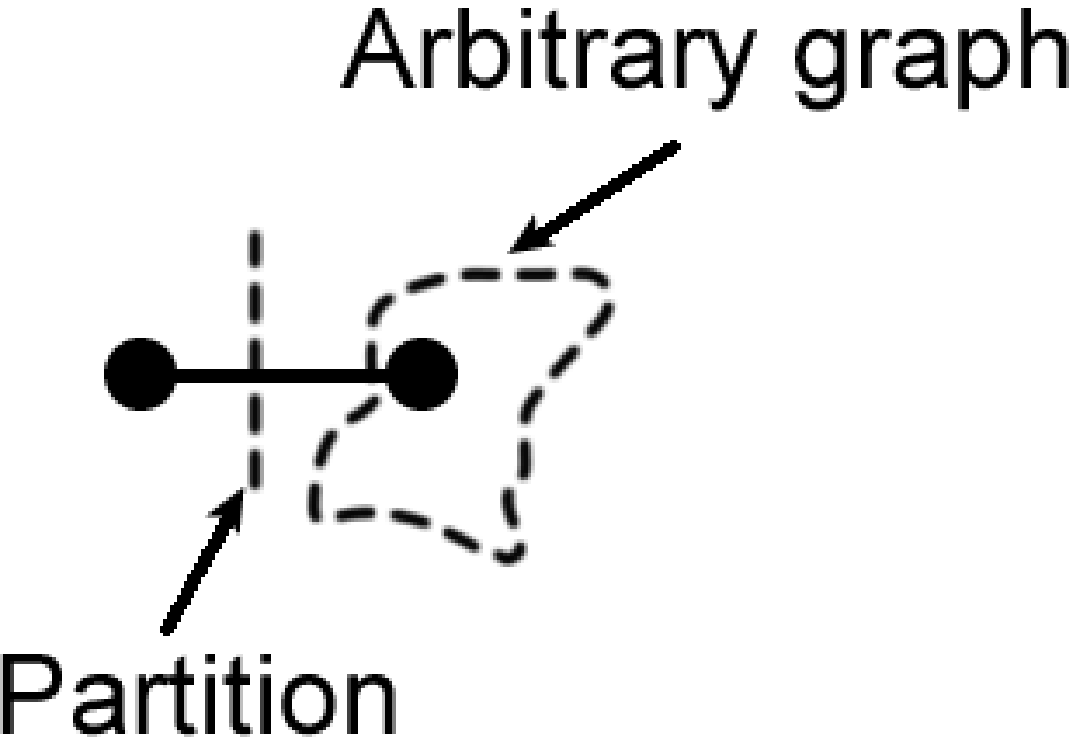}
\end{center}
All other graphs must contain a loop.
If we take the smallest loop in the graph, then the qubits in this loop
already have at least two neighbors. We start by taking the triangle, a loop
of 3 qubits. With no additional connections, this state is non-LR.
\begin{center}
\includegraphics[width=0.04\textwidth]{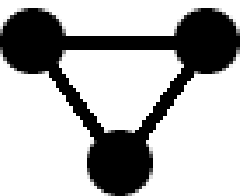}
\end{center}
If we add one extra qubit, then by connecting it to two or three qubits in
the triangle, the state is non-LR.
\begin{center}
\includegraphics[width=0.09\textwidth]{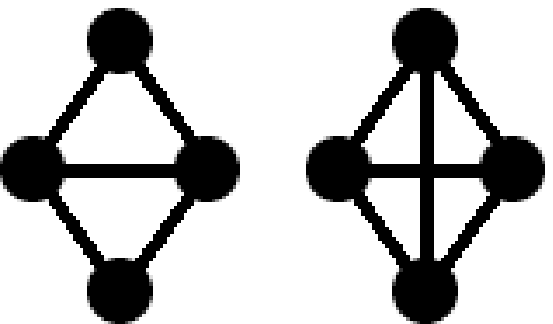}
\end{center}
When the extra qubit is only connected to two qubits, there are two further
links that could be added to any arbitrary structure. If they are not
connected to the same qubits, these two links provide an LR partition.
\begin{center}
\includegraphics[width=0.3\textwidth]{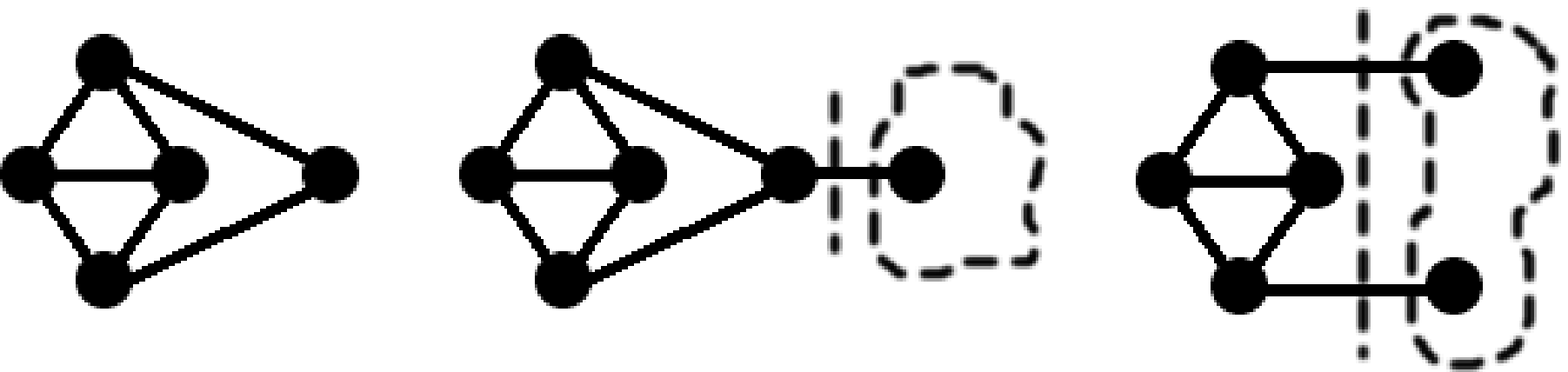}
\end{center}
The only remaining structure is where all three links from the triangle are
connected to some arbitrary graph, but are not incident on the same qubit.
These extra links must provide an LR partition.

We now continue this argument to loops of 4 qubits. Any
structure which we add that generates a triangle has, of course, already
been dealt with. This only leaves four examples which are non-LR.
\begin{center}
\includegraphics[width=0.26\textwidth]{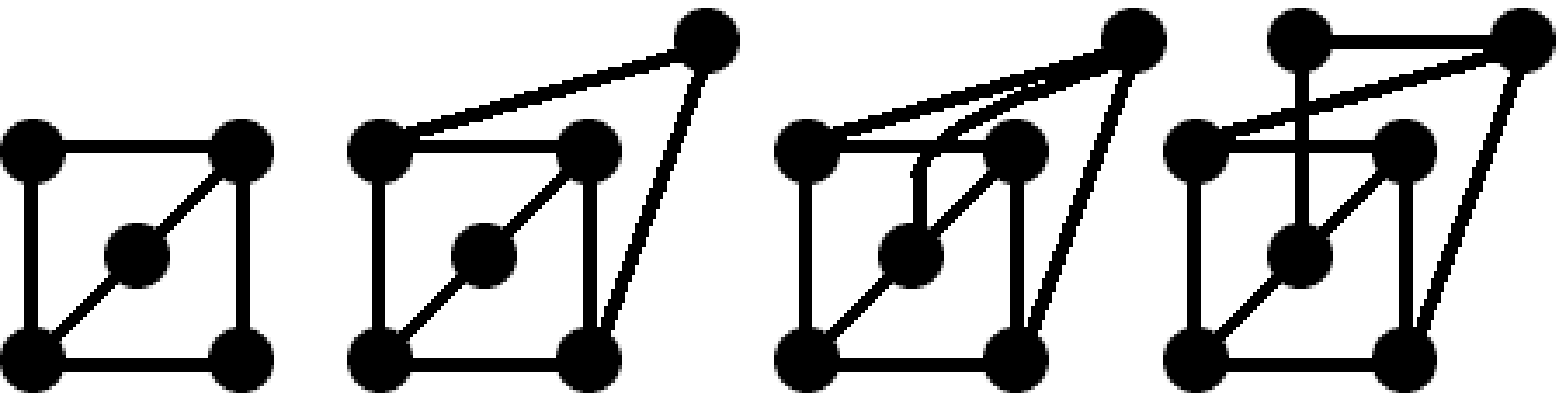}
\end{center}
Finally, for larger loops, there is no way of creating a graph that is
non-LR without forming smaller loops first. Consequently, for graphs of
maximum degree 3, there are only 8 graphs which are non-LR. It can be
verified that the latter are locally equivalent to LR states.

\subsection{Size of LR Class} \label{sec:size}

\begin{table}
\begin{tabular}{|c|c|c|}
\hline
N & LR states & All graph states \\
\hline
3 & 7 & 8 \\
4 & 53 & 64 \\
5 & 788 & 1024 \\
6 & 22204 & 32768 \\
7 & 1148781 & 2097152 \\
\hline
\end{tabular}
\caption{Comparison of the number of LR states of $N$ qubits
and the total number of graph states. Note that we have included the
completely separable state in the set of LR states.}\label{tab:LR}
\end{table}

It is important to answer the question of how large the set of LR states is.
The set of all graph states of $N$ qubits consists of $2^{\binom{N}{2}}$
elements, including all possible isomorphisms and graphs which can be
separated into two or more unconnected sub-graphs. We can easily generate LR
states from the $(N-1)$-qubit graphs by adding an extra qubit and connecting
it to any single qubit from the previous graph. This state must be LR
because the qubit that we have just added constitutes one such partitioning.
There are at least $(N-1)2^{\binom{N-1}{2}}$ such states. Despite being a
small fraction ($\sim N/2^{N-1}$) of all graph states, they certainly form a
significant class in their own right. Moreover, there are further examples
to be added into the class of LR graph states, but accurate enumeration is a
combinatorial challenge. In Table \ref{tab:LR} we present the number of LR
states for graphs of up to 7 qubits.

Now that we have provided a lower bound, is it possible to either tighten
this bound, or give an upper bound? Let us consider all possible graph
states of $N$ qubits. We shall select a specific partition of $q<N-q$
qubits. The probability, $p_q$, that local reconstruction is possible across
this boundary is given by
$$
p_q=\sum_{b=0}^q\frac{\binom{q}{b}\binom{N-q}{b}b!}{2^{q(N-q)}}.
$$
This is a result of requiring that there are $q(N-q)$ possible bonds across
the partition, which could either be bonded or not. Of these $2^{q(N-q)}$
combinations, only those with no more than a single bond from each qubit
make the state locally reconstructible. For $b$ bonds across the
partition, we have to choose them from $q$ on one side and $N-q$ on the
other side. Finally, the ordering of the choice on one side of the partition
is important, hence the $b!$. This can be expressed in terms of the
confluent hypergeometric function,
$$
p_q=2^{q(q-N)}(-1+(-1)^qU(-q,1-2q+N,-1)).
$$

There are $\binom{N}{q}$ ways that we could have chosen a partition
of $q$ qubits. Each of these has the same probability of giving local
reconstructibility, but we must make sure we do not over count the cases
where there is more than one partitioning for the same graph state,
$$
P_q=-\sum_{n=1}^{\binom{N}{q}}(-1)^n\binom{\binom{N}{q}}{n}p_q^n
=1-(1-p_q)^{\binom{N}{q}}.
$$
Similarly, we need not be restricted to a specific $q$, but must avoid
over-counting,
$$
P=\sum_{i=1}^{\lfloor N/2\rfloor}P_i-\sum_{j<i}P_iP_j+\ldots.
$$
To simplify this expression, we can take the smallest (largest) value of
$P_q$ and assume that all $P_q$ have this value, thereby lower (upper)
bounding $P$.
$$
P\geq-\sum_{n=1}^{\lfloor N/2\rfloor}\binom{N/2}{n}(-1)^nP_\text{min}^n=
1-(1-P_\text{min})^{N/2}.
$$
The bounds will occur for $q=1$ and $q=N/2$. In the case of $q=1$, $p_q=
(N-1)/2^{N-1}$ and hence the fraction of LR states is estimated to be
$N^3/2^N$. To see that this is an upper bound, we write that $p_q\sim
2^{q(q-N)}O(N^q)$ and $P\approx p_q\binom{N}{q}\frac{N}{2}$, having assumed
that a particular value of $q$ is going to give the required bound. The
ratio for successive values of $q$ is therefore given by
$$
\frac{P|_q}{P|_{q-1}}=\frac{N-q}{q}\cdot O(N)\cdot 2^{2q-N-1}.
$$
Hence, each successive value of $P$ must be smaller given the overpowering
nature of the exponential $2^{-N}$. Therefore the bound which we have just
derived is an upper bound. Combining this with our existing lower bound,
$$
\frac{N-1}{2^{N-1}}\leq \frac{|LR_N|}{|G_N|}\leq\frac{N^3}{2^N}.
$$

\section{The Performance of the Divide and Rebuild Protocol}
\label{sec:derivation}

In the body of the paper, we focused on calculating upper bounds for certain
types of noise, and comparing them to a numerical analysis of the performance of
the GMPP. We have also analyzed the DRPP in the case of local-$Z$ noise,
when we can show that it is optimal. In general, we do not expect this
protocol to be optimal, but it is still useful because we can analyze its
performance, and use it to place a lower bound on the performance of any
optimal protocol.

\subsection{Maximally Depolarizing Noise}

We can readily show that the DRPP does not adapt well in the presence of
some correlated noise (of course, $X$ and $Y$-errors can be represented as
correlated $Z$-errors, so we have optimality in some special cases). For
example, we can take the case of the maximally depolarized state of $N$
qubits,
$$
\rho_N=\frac{\identity+x\proj{\psi_G}}{2^N+x}.
$$
and consider $\ket{\psi_G}$ to be the $N$-qubit linear cluster state. When
we perform measurements on this state, we reduce it from
$\rho_N(x)\rightarrow\rho_{N-1}(x/2)\rightarrow\rho_2(x/2^{N-2})$. The
bipartite state can be purified if
$$
\frac{1+\frac{x}{2^{N-2}}}{4+\frac{x}{2^{N-2}}}>\half
$$
and hence the fidelity goes as
$$
\bra{\psi_G}\rho\ket{\psi_G}>\frac{1}{3}+\frac{1}{3\cdot2^{N-1}}
$$
which tends to a fixed value of $1/3$, whereas genuine multipartite
purification protocols can purify states exponentially decreasing
fidelities~\cite{aschauer:04}.

\subsection{Local Depolarizing Noise}

We would also like to demonstrate the performance of the DRPP when we do not
know what the type of local error is. We proceed by assuming that the noise
is the most destructive type of local noise considered here. This is simply
$Z$-noise, because these errors form a complete basis for the state, whereas
other types of error need not. Errors occur with
probability $p$, with an equal likelihood of them being either $X$, $Y$ or
$Z$. Making a $Z$-measurement hence causes the propagation of an error to
another qubit in $2/3$ of the cases. The simplicity of the protocol,
however, continues to help us, enabling the calculation of the critical
error probability, $p_\text{crit}$ assuming that the graph has a maximum
degree of $D_G$. Note that it is necessary to assume that the graph is
two-colorable, which means that an error can only propagate to one of the
two qubits in the Bell pair to be purified.

Consider a single Bell-pair, which is the one which we aim to measure
towards and purify. Each qubit is attached to $D_G-1$ other qubits in the
graph. We will be able to purify the state if the probability of error on
the Bell pair, after measurement of the other qubits, is better than \half.
The probability of an error occurring on a single qubit is $p$, and it is
equally likely to be $X$, $Y$ or $Z$. Let $q$ be the probability that all
the qubits attached to one of the pair give no errors on the qubit they are
attached to after $Z$-measurements. This is caused by no errors occurring,
$Z$ errors occurring (which do not get transmitted) or pairs of $X$ or $Y$
errors, which cancel when transmitted to the final qubit.
\begin{eqnarray}
q&=&\sum_{n=0}^{\lfloor\half(D_G-1)\rfloor}\binom{D_G-1}{2n}
\left(\frac{2p}{3}\right)^{2n}  \nonumber\\
&\times&\sum_{m=0}^{D_G-1-2n}\binom{D_G-1-2n}{m}(1-p)^{D_G-1-2n-m}
\left(\frac{p}{3}\right)^m    \nonumber\\
&=&\half\left(1+(1-4p/3)^{D_G-1}\right) \nonumber
\end{eqnarray}
If $g(n)$ is the probability that $n$ errors occur on the Bell pair due to
its local errors (i.e. neglecting the connected qubits), then after
measurements there is no error with a probability
$$
g(0)q^2+g(1)q(1-q)+g(2)(1-q)^2.
$$
No errors occur on the Bell pair if  no errors truly occurred, $(1-p)^2$, or
pairs of errors cancel ($YY$, $ZX$, $XZ$), $p^2/3$. Hence,
$g(0)=(1-p)^2+p^2/3$. Similarly, we find that $g(2)=\half g(1)=1-g(0)/3$.
Substitution of these yields the polynomial
$$
x^{2D_G}+2x^{D_G+1}-1=0
$$
where $x=1-4p/3$ and $p$ is the critical error rate below which this
protocol will perfectly purify the state. In the case of the Steane code,
$D_G=3$, and hence $p_\text{crit}=0.16$. As stated in the body of the paper,
existing multipartite purification protocols improve upon this probability.

\subsection{Fully Connected Graph States} \label{sec:nonLR}

As an example of the application of the DRPP, we will examine the
purification of fully-connected graph states. Note that these graphs are not
LR, although they are locally equivalent to GHZ states, and hence have been
indirectly included in previous discussions. We shall start by considering
the triangular configuration of Fig.~\ref{fig:graphs}(c).

Given the small size of the triangle, examination of the different local
error combinations is tractable. There are $3^3=27$ different combinations
of local errors. Of these, 12 can be turned into independent local $Z$-noise
on the 3-qubit chain using the equivalence of graphs under local unitaries. These are the combinations $XXY$, $XYZ$
and $ZZY$ and their permutations. Two further cases of particular interest
are local $X$-noise and local $Y$-noise. Our protocol of measuring a single
qubit (in the $X$-basis in this case) can tolerate an error probability of
$p<\half$ for $X$-noise. This is because pairs of errors obey identities
such as $X_2X_3=X_1$, and hence an $X$-measurement on qubit 1 commutes with
these errors.

The ultimate realization of this is in the case of independent local
$Y$-noise, since a $Y$-error is the same as correlated $Z$-errors on all 3
qubits. Hence, all the $Y$-errors are the same, and a single $Y$-measurement
would eliminate all of them. However, a single $Y$-measurement leaves all
the other qubits in a separable state -- all the bonds are broken, not just
those connected to the qubit that is measured. Instead, we must perform some
other measurement. A $Z$-measurement, for example, allows purification for
all errors since one of the two diagonal elements is always larger than
$\half$.

The results derived for a triangular graph for both local $X$-noise and
local $Y$-noise hold for all fully connected graphs,
$$
\ket{G_N}=\frac{1}{\sqrt{2^N}}\sum_{i=0}^{2^N-1}(-1)^{\binom{w_i}{2}}\ket{i},
$$
where $w_i$ is the binary weight of the number $i$. In the case of
$Y$-noise, a single $Y$-error is the same as correlated $Z$-errors across
all qubits, so all $Y$-errors are the same, and the noisy state only has two
diagonal elements, the first of which is the probability that an even number
of errors occurred,
\begin{eqnarray}
f&=&\sum_{n=0}^{\lfloor N/2\rfloor}(1-p)^{N-2n}p^{2n}\binom{N}{2n}
\nonumber\\
&=&\half\left((1-2p)^N+1\right) \nonumber
\end{eqnarray}
Upon performing a $Z$-measurement on one of the qubits, we are left with the
state $\ket{G_{N-1}}$ and the diagonal elements must be the same as they
were before. This can be verified by writing
\begin{equation}
\ket{G_N}=\frac{1}{\sqrt{2^N}}
\sum_{j=0}^{2^{N-1}-1}(\ket{0}+
(-1)^{w_j}\ket{1})(-1)^{\binom{w_j}{2}}\ket{j}.    \label{eqn:GN}
\end{equation}
Note that $(-1)^{w_j}\ket{j}=Z^{\otimes N}\ket{j}$ if $j$ is an $N$-bit
binary string, which means that if we get the $\proj{1}$ result, the
corrective unitary is a $Z$-rotation on every qubit. This reduction to fully
connected graphs continues to the case of $\ket{G_2}$, which we know can
always be purified if it only has two diagonal elements since one element is
always greater than $\half$.

In the case of $X$-noise, we must verify that an $X$-measurement on a single
qubit reduces $\ket{G_N}$ to $\ket{G_{N-1}}$. After applying $\proj{+}$ to
the state in Eqn.~(\ref{eqn:GN}), we are left with a sum over binary
strings of even weight. To convert this into $\ket{G_{N-1}}$, we apply a
Hadamard on one qubit, and $Z$-rotations on all the other qubits. Since the
independent $X$ errors commute with this measurement process, this will
eventually reduce to the triangle, which we have already solved, and hence
$p<\half$ is the criterion for purification. Note, however, that when
reducing to a two-qubit state, our density matrix has 3 diagonal elements,
not 2 as in the case of $Y$ errors, so purification at $p>\half$ is not
possible with this protocol.

\section{Fixed Points of Bipartite Purification} \label{sec:fixed_derive}

In \cite{alber:01}, a bipartite purification protocol was proposed for which
subsequent analysis \cite{delgado:03} showed that its behavior is very
similar to that of the GMPP, with the added advantage that we can calculate
its performance. As an example, let us consider maximally depolarizing noise
applied to a linear state of $N$ qubits. For simplicity, and consistency
with \cite{delgado:03}, we shall define $D=2^{N/2}$, assuming that there are
$N/2$ qubits either side of the bipartite split. The diagonal elements of
the initial density matrix, $\lambda_{kj}^{(0)}$ indicate the errors $k$ and
$j$ either side of the partition. After $n$ iterations of the protocol, the
unnormalized outcome is
$$
\lambda_{kj}^{(n)}=\sum_{k'=0}^{D-1}e^{-\frac{2\pi ikk'}{D}}
\left[\sum_{k''=0}^{D-1}e^{-\frac{2\pi ik''k'}{D}}\lambda_{k''j}^{(0)}
\right]^{(2^n)}.
$$
For maximally depolarizing noise,
$$\lambda_{kj}^{(0)}=\frac{1}{D^2+x}+\delta_j\delta_k\frac{x}{D^2+x}.$$
Following this through and renormalizing, we find that
$$
\lambda_{00}^{(n)}=\frac{(x+D)^{2^n}+(D-1)x^{2^n}}
{D\left[(x+D)^{2^n}+(D-1)D^{2^n}\right]}.
$$
For $x=D$, $\lambda_{00}^{(n)}$ is independent of $n$, and hence we have
found a fixed point of the protocol. The implication is that there is a
critical fidelity of $1/2^{N/2}$, below which the maximum achievable
fidelity is $1/2^{N/2}$, and above which complete purification is possible.

\end{document}